\documentclass[10pt,a4paper,oneside,english]{elsart}
\usepackage[T1]{fontenc}
\usepackage[latin1]{inputenc}
\usepackage{float}
\usepackage{graphicx}
\usepackage{amssymb}

\usepackage{babel}
\usepackage{epsfig}
\makeatother
\begin{document}
\begin{frontmatter}

\title{Study of the Correlations Between Stocks of Different Markets}

\author[TCD_Physics]{Ricardo Coelho},
\ead{rjcoelho@tcd.ie}
\author[TCD_Physics]{Peter Richmond},
\author[TCD_Physics]{Stefan Hutzler}
\author[TCD_Bus]{\and Brian Lucey}

\address[TCD_Physics]{School of Physics, Trinity College Dublin, Dublin 2, Ireland}
\address[TCD_Bus]{Institute for International Integration Studies and School of Business, Trinity College Dublin, Dublin 2, Ireland}

\begin{abstract}
We study correlations of a set of stocks selected from both the New York and London stock exchanges. Results are displayed using both Random Matrix Theory approach and the graphical visualisation of the Minimal Spanning Tree. For the set of stocks we study, cross correlations between markets do not mix the markets significantly. Geographical differences seem to dominate the output of a Random Matrix analysis. Only at the level of the third highest eigenvector do we see an effect of New York on the London data with the emergence of some common sectors with the larger eigenvectors in London and New York. The Minimal Spanning Trees show the broad separation of the markets as reflected in the second eigenvector of the Random Matrix analysis. However more detail is difficult to discern from the Minimal Spanning Trees analysis.
\end{abstract}

\begin{keyword}
Econophysics; minimal spanning tree; random matrix theory.

\PACS{89.65.Gh}
\end{keyword}
\end{frontmatter}

\section{Introduction}

The relation between the returns of two different companies can be quantified by computing the correlation between the time series of prices of both companies. For a portfolio of stocks this leads to a correlation matrix. The Minimal Spanning Tree approach uses some of the information contained in this matrix to obtain a graphical representation of the correlations. Many empirical studies have shown that within the constructed tree, stocks cluster in groups of the same industrial sector \cite{{Mantegna_EPJB11_193_1999},{Onnela_PRE68_056110_2003},{Coelho_PhysA_373_615}}. Minimal Spanning Trees studies of different indices of markets have shown that these cluster according to geographical location \cite{{Bonanno_PRE62_7615_2000},{Coelho_PhysA_376_455}}. A different approach in dealing with the correlation matrix consists of a numerical computation of its eigensystem \cite{{Gopikrishnan_PRE_64_035106_2001},{Plerou_PRE_65_066126_2002},{Utsugi_PRE_70_026110_2004},{Kim_PRE72_046133_2005}}. The eigenvectors that correspond to the highest eigenvalues show segregation for stocks of different industrial sectors.

In this paper we compute correlations between stocks on the London (LON) and New York (NYSE) markets. In selecting the set of stocks we have here chosen $939$ large company stocks across all the sectors as defined by the ICB classification \cite{ICBClassification}. However we have left out of our set of stocks those stocks that are quoted on both the LON and NYSE markets. The data for each stock is the daily closing price in USD for the $3127$ trading days from $30^{th}$ December $1994$ to $1^{st}$ January $2007$.
The results of both Random Matrix analysis and Minimal Spanning Tree show that LON and NYSE markets remain separated. However in the second and third largest eigenvectors of the correlation matrix it can be seen that NYSE does affect the LON market with cross correlations enhancing certain sectors.

In section 2 we review the methodology. The results for the separate markets, London and New York are then shown in section 3. Section 4 shows how these are modified when cross correlations between NYSE and LON are introduced. Finally the conclusions are presented in section 5.

\section{Definitions}

Our study is based on the assumption that the returns of the stock price carry more information than random noise. To check this, we compute the correlation between returns of $N$ stock prices and analyse the correlation matrix. The correlation coefficient, $\rho_{ij}$ between stocks $i$ and $j$ is given by:
\begin{equation}
\rho_{ij}=\frac{\langle{\bf R}_{i}{\bf R}_{j}\rangle-\langle{\bf R}_{i}\rangle\langle{\bf R}_{j}\rangle}{\sqrt{\left(\langle{\bf R}_{i}^{2}\rangle-\langle{\bf R}_{i}\rangle^{2}\right)\left(\langle{\bf R}_{j}^{2}\rangle-\langle{\bf R}_{j}\rangle^{2}\right)}}
\label{CorrelCoefEq}
\end{equation}
where ${\bf R}_{i}$ is the vector of the time series of log-returns, $R_{i}(t)=\ln P_{i}(t)-\ln P_{i}(t-1)$ and $P_{i}(t)$ is the price of stock $i$ at time $t$. The notation $\langle\cdots\rangle$ means an average over time $\frac{1}{T}\sum_{t'=t}^{t+T-1}\cdots$, where $t$ is the beginning of the series and $T$ is the length of the time series.
We can normalise the time series of returns for each stock by subtracting the mean and dividing by the standard deviation:
\begin{equation}
  \tilde{{\bf R}}_{i}=\frac{{\bf R}_{i}-<{\bf R}_{i}>}{\sqrt{\langle{\bf R}_{i}^{2}\rangle-\langle{\bf R}_{i}\rangle^{2}}}
\label{eq_normal}
\end{equation}
The correlation coefficient is then given by: $\rho_{ij} = \langle\tilde{{\bf R}}_{i} \tilde{{\bf R}}_{j}\rangle$ \cite{Mantegna_Book}.
This coefficient can vary between $-1\leq\rho_{ij}\leq1$, where $-1$ means completely anti-correlated stocks and $+1$ completely correlated stocks. If $\rho_{ij}=0$ the stocks $i$ and $j$ are uncorrelated. The coefficients form a symmetric $N\times N$ matrix with diagonal elements equal to unity.
The correlation matrix with elements $\rho_{ij}$ can be represented as:
\begin{equation}
{\sf C}=\frac{1}{T}{\sf G} {\sf G}^T
\end{equation}
where ${\sf G}$ is an $N\times T$ matrix with elements $\tilde{R}_{i}(t)$ and ${\sf G}^T$ denotes the transpose of ${\sf G}$.

\subsection{Random Matrix Theory}
Important information about the financial data is obtained by studying the eigensystem of the correlation matrix. In particular the spectrum of eigenvalues differs markedly from the one for random matrices \cite{{Gopikrishnan_PRE_64_035106_2001},{Plerou_PRE_65_066126_2002}}. A random matrix is defined by \cite{Sengupta_PRE_60_3389}:
\begin{equation}
{\sf C'}=\frac{1}{T}{\sf G}'{\sf G}'^T
\end{equation}
where ${\sf G}'$ is a $N\times T$ matrix with columns of time series with zero mean and unit variance, that are uncorrelated. The spectrum of eigenvalues can be calculated analytically. In the limit $N\rightarrow \infty$ and $T\rightarrow \infty$, with $Q=T/N$ fixed, and bigger than $1$, the probability density function of eigenvalues of the random matrix is given by:
\begin{equation}
P_{RM}(\lambda)=\frac{Q}{2\pi}\frac{\sqrt{(\lambda_{max}-\lambda)(\lambda-\lambda_{min})}}{\lambda} .
\label{RM_spectrum}
\end{equation}
Here 
\begin{equation}
\lambda_{min}^{max}=\left(1\pm \frac{1}{\sqrt{Q}}\right)^2
\label{RMT_lambda}
\end{equation}
limits the interval where the probability density function is different from zero. The eigenvalues outside these limits contain information about the correlations of the time series studied as will be shown below. This information is contained in the elements of the eigenvectors that belong to each of these eigenvalues. 

Each eigenvector contains $N$ elements, each of them related to one stock. When we study a portfolio of stocks from just one market, using the ICB classification \cite{ICBClassification}, we group each stock in its industrial sector. For the study of stocks from more than one market we divide each industrial group in markets. We compute a value for each market/industrial sector group, where we calculate the mean:
\begin{equation}
  I_{S_k}^{M_j} = \frac{1}{N_{S_k}^{M_j}}\sum_{i\in M_j, i\in S_k} I_i 
\end{equation}
where $I_i$ is the element $i^{th}$ of the eigenvector, $S_k$ represent the sector $k$ ($k=1,\cdot \cdot \cdot$) and $N_{S_k}^{M_j}$ is the number of stocks that belong to sector $k$ and market $j$. This new quantity give us some information about each sector of each market, instead of the normal information of each stock.

\subsection{Minimal Spanning Tree}

Another way to study the correlation of stocks is to create a matrix of distances between stocks from the correlation coefficients. With this matrix of distances we can create a tree where nodes are stocks and links are the distance between the stocks. If two stocks are highly correlated, the distance between them is small. The tree that we use to study these properties is the Minimal Spanning Tree (MST).
The metric distance, introduced by Mantegna \cite{Mantegna_EPJB11_193_1999}, is determined from the Euclidean distance between vectors, $d_{ij}=|\tilde{{\bf R}}_{i}-\tilde{{\bf R}}_{j}|$. Because $|\tilde{{\bf R}}_{i}|=1$ (see eq. \ref{eq_normal}), it follows that:
\begin{equation}
d_{ij}^{2}=|\tilde{{\bf R}}_{i}-\tilde{{\bf R}}_{j}|^{2}=|\tilde{{\bf R}}_{i}|^{2}+|\tilde{{\bf R}}_{j}|^{2}-2\tilde{{\bf R}}_{i}\cdot\tilde{{\bf R}}_{j}=2-2\rho_{ij}
\end{equation}
The relation between the distance of two stocks and their correlation coefficient is thus given by:
\begin{equation}
d_{ij}=\sqrt{2(1-\rho_{ij})}\label{Distance}\end{equation}
This distance varies between $0\leq d_{ij}\leq2$. Following the procedure of Mantegna \cite{Mantegna_EPJB11_193_1999}, this distance matrix is now used to construct a network which contains the essential information of the market. 

This network (MST) has $N-1$ links connecting $N$ nodes. The nodes represent stocks and the links are chosen such that the sum of all distances (normalised tree length) is minimal. We perform this computation using Prim's algorithm \cite{Prim}.
The main idea for using MST, apart of the visualisation of links between companies, is to filter data. From the $N\times (N-1)/2$ correlation coefficients we are only left with $N-1$ points, which represent the most important information of the correlation matrix.

\section{Data from two different markets}

\begin{figure}[H]
  \begin{center}
    \epsfysize=80mm
    \epsffile{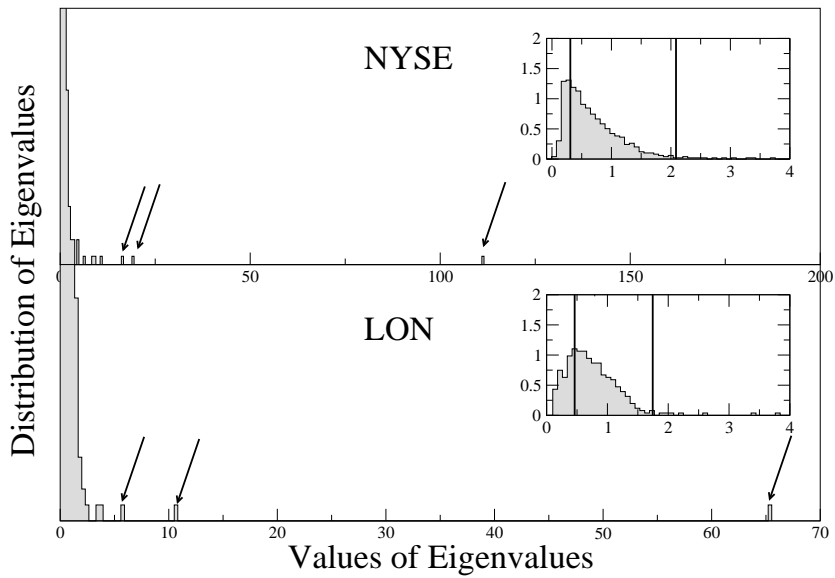}
    \caption{Spectrum of eigenvalues for two different portfolios: a portfolio of $617$ stocks from NYSE (top); a portfolio of $322$ stocks from LON (bottom). The vertical lines, in the inset figures, show the limits $\lambda_{min}^{max}$ (eq. \ref{RMT_lambda}). The arrows show the three highest eigenvalues for each market that we study more carefully in this paper.}
    \label{figure_1}
  \end{center}
\end{figure}

The distributions of the eigenvalues of the correlation matrix for the markets of NYSE and LON are shown in Figure \ref{figure_1}. The largest eigenvalue for each market seems to depend on the size of the portfolio or probably in the correlation of the stocks in the portfolio.

\begin{figure}[H]
  \centering
  \begin{tabular}{cc}
    \begin{minipage}{7.cm}
      \includegraphics[width=0.9\textwidth]{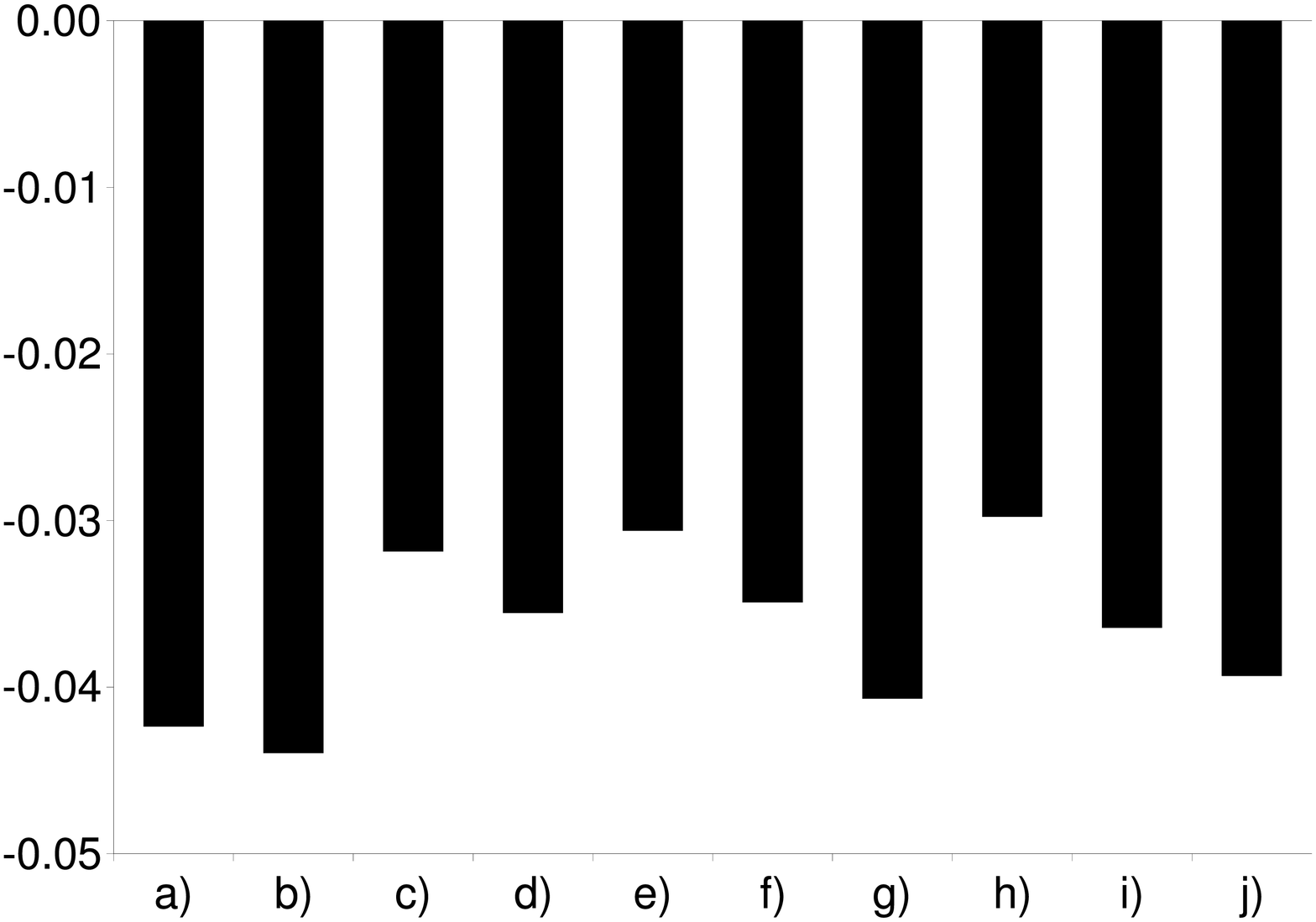}
      \begin{center}
	NYSE
      \end{center}
    \end{minipage}
    \begin{minipage}{7.cm}
      \includegraphics[width=0.9\textwidth]{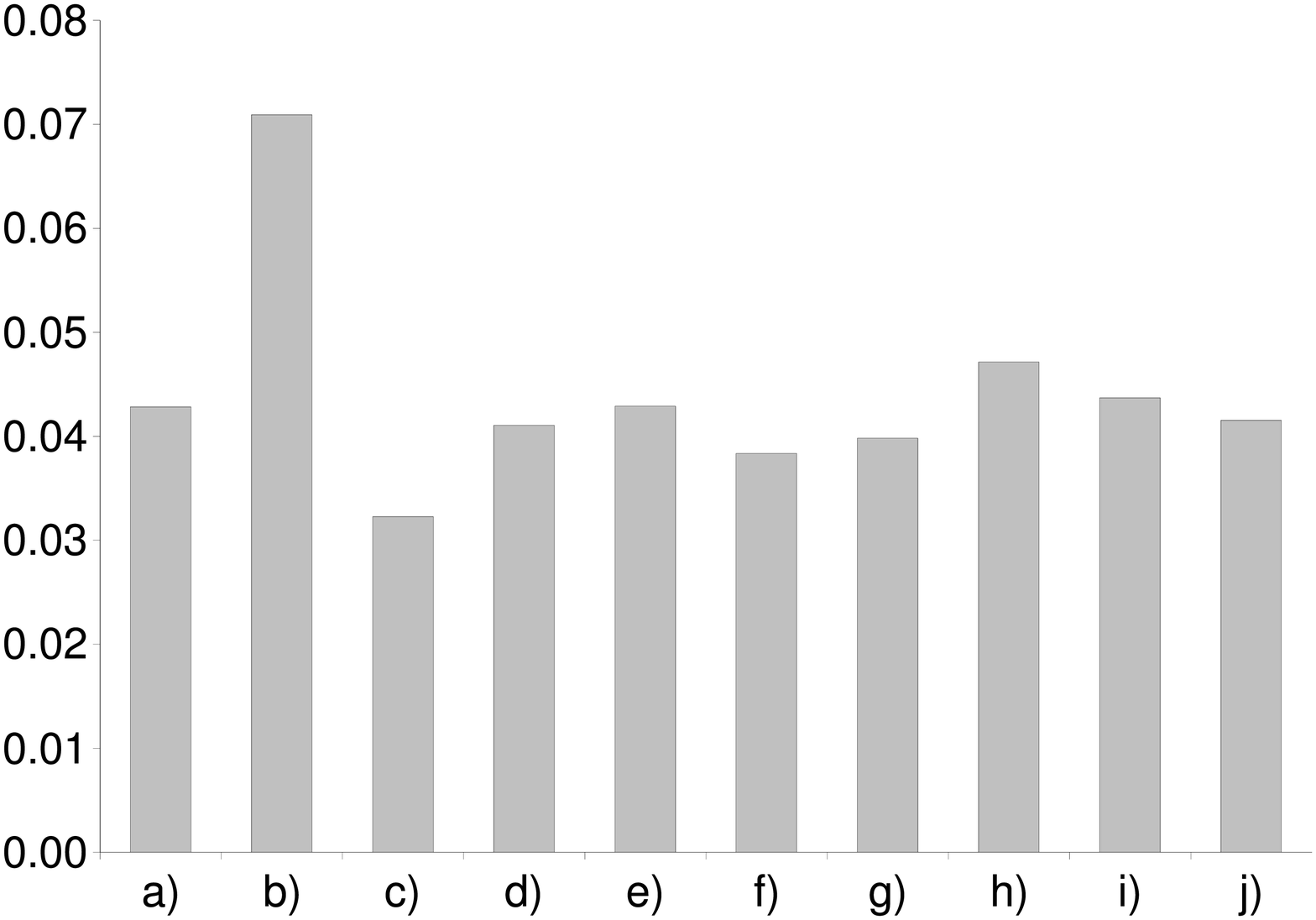}
      \begin{center}
	LON
      \end{center}
    \end{minipage}
  \end{tabular}
  \caption{Eigenvector elements of the highest eigenvalue for two different markets: NYSE and LON. In the x axis we have the group of elements that belong to a industrial sector: a) industrials; b) financials; c) health care; d) technology; e) oil and gas; f) utilities; g) basic materials; h) telecommunications; i) consumer goods; j) consumer services. All industrials sectors of a market are of the same sign. Note that the different signs for NYSE and LON data are irrelevant since eigenvectors remain eigenvectors when multiplied by $(-1)$.}
  \label{figure_2}
\end{figure}

\begin{figure}[H]
  \centering
  \begin{tabular}{cc}
    \begin{minipage}{7.cm}
      \includegraphics[width=0.9\textwidth]{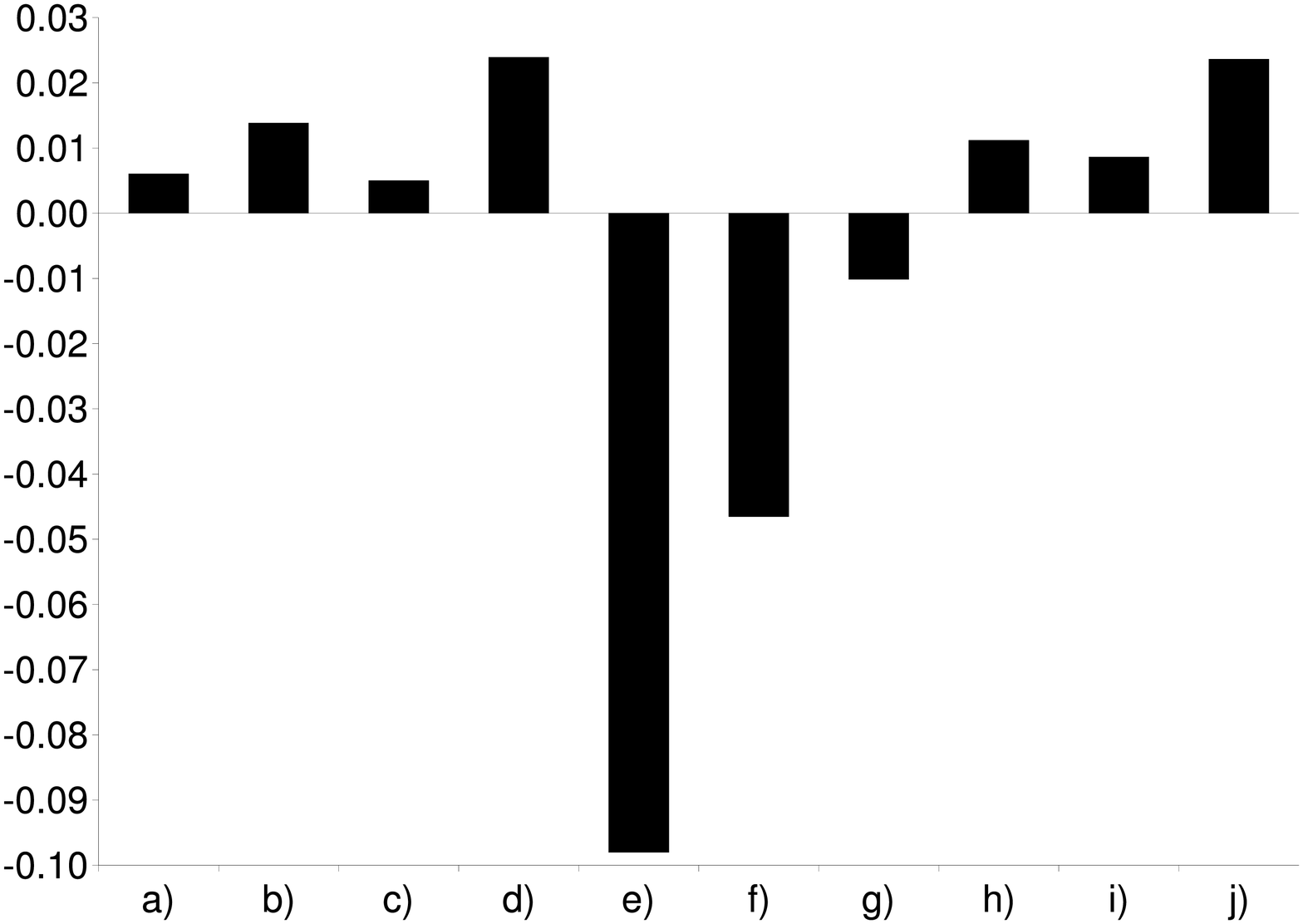}
      \begin{center}
	NYSE
      \end{center}
    \end{minipage}
    \begin{minipage}{7.cm}
      \includegraphics[width=0.9\textwidth]{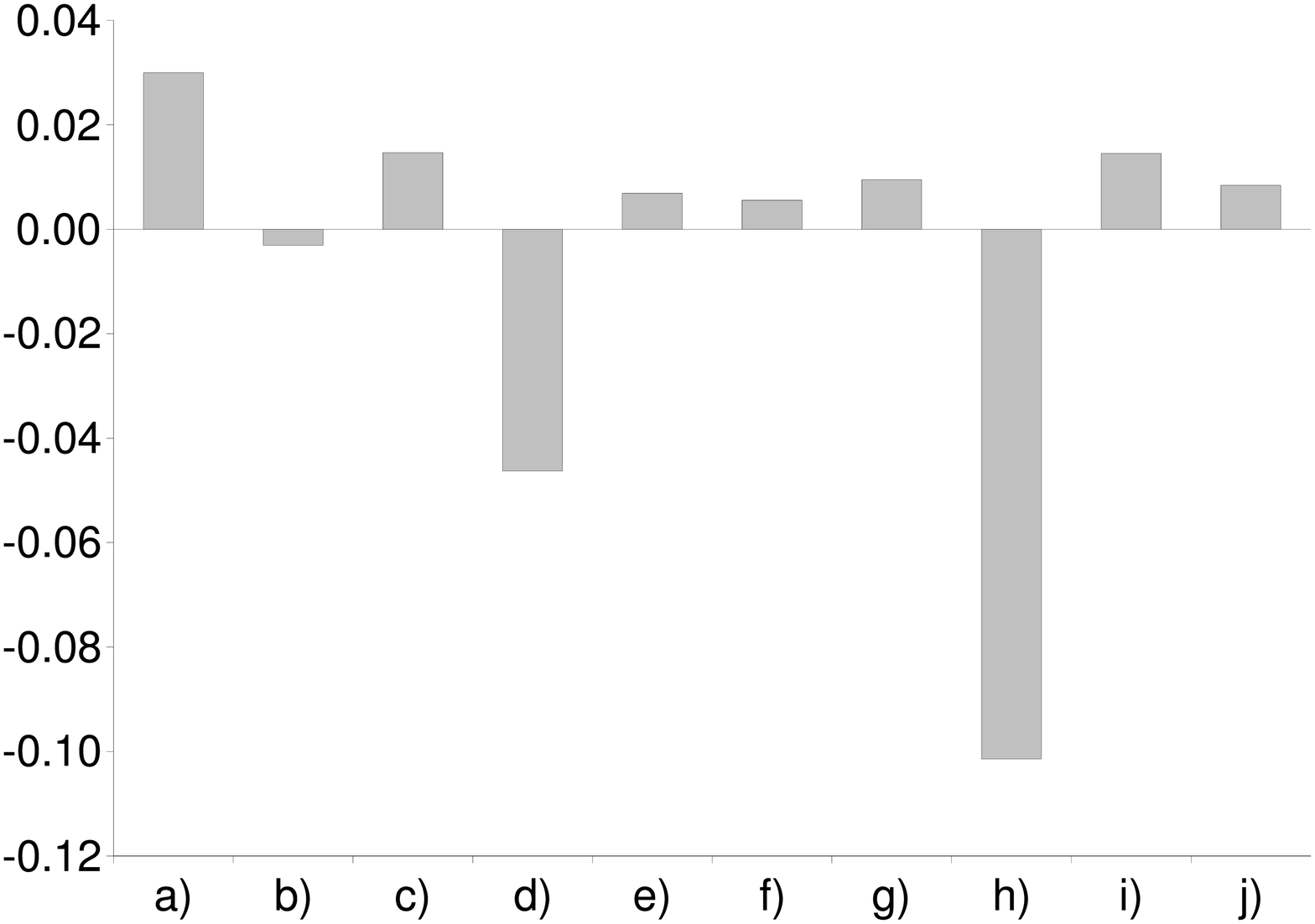}
      \begin{center}
	LON
      \end{center}
    \end{minipage}
  \end{tabular}
  \caption{Eigenvector elements of the $2^{nd}$ highest eigenvalue for the markets NYSE and LON. Oil and gas and utilities for NYSE and telecommunications and technology for LON are the largest contributions.}
  \label{figure_3}
\end{figure}

\begin{figure}[H]
  \centering
  \begin{tabular}{cc}
    \begin{minipage}{7.cm}
      \includegraphics[width=0.9\textwidth]{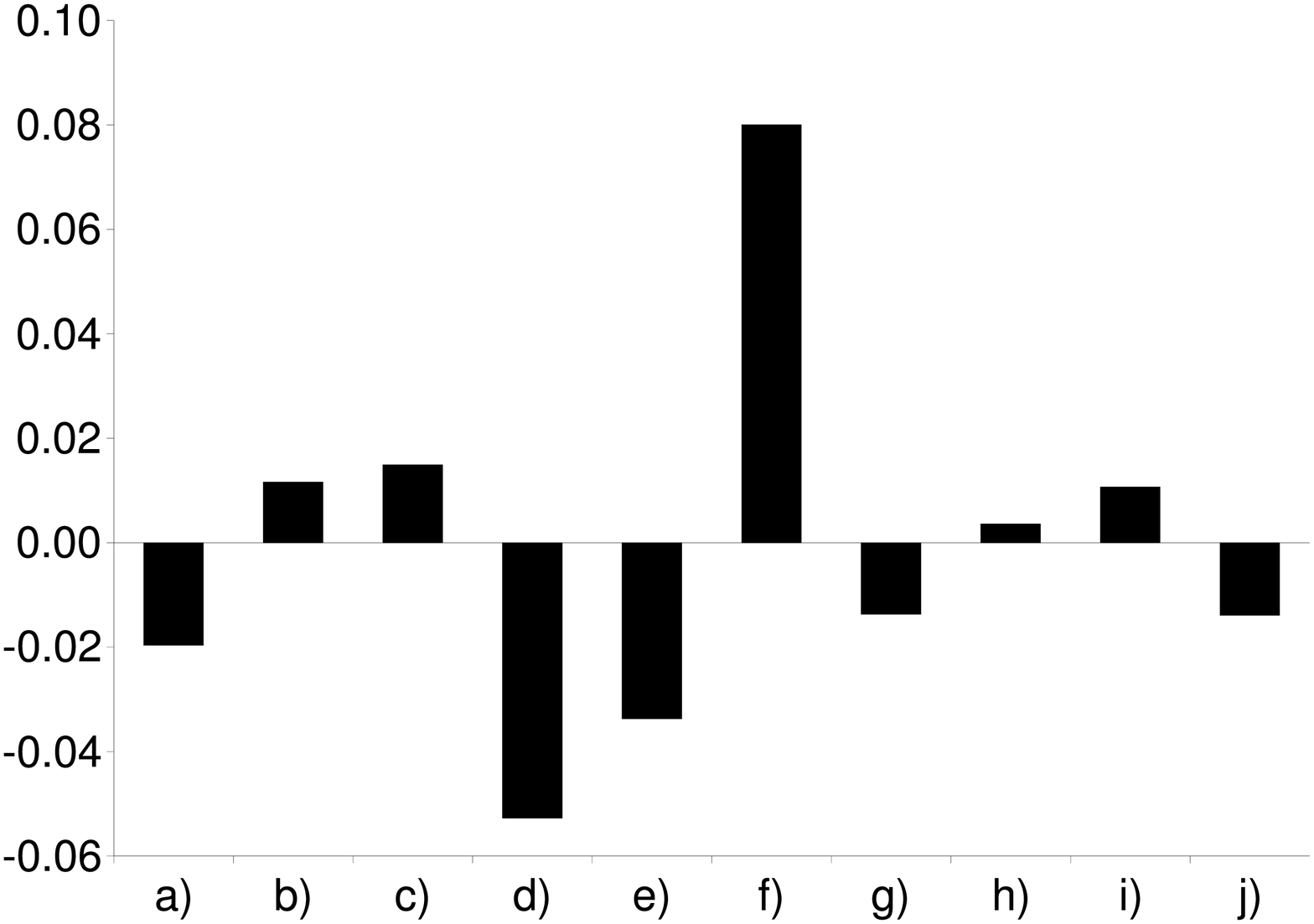}
      \begin{center}
	NYSE
      \end{center}
    \end{minipage}
    \begin{minipage}{7.cm}
      \includegraphics[width=0.9\textwidth]{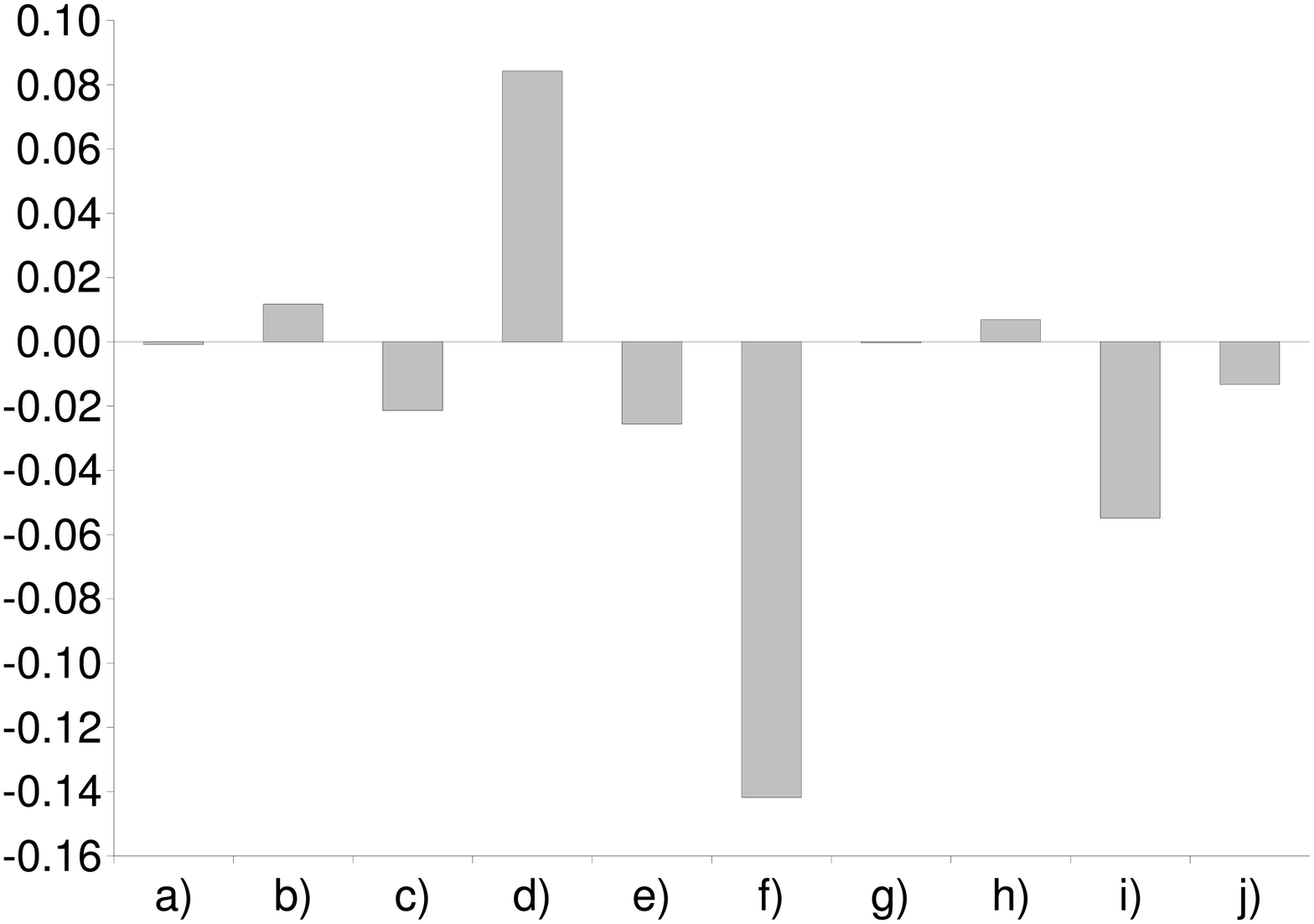}
      \begin{center}
	LON
      \end{center}
    \end{minipage}
  \end{tabular}
  \caption{Eigenvector elements of the $3^{rd}$ highest eigenvalue for the markets NYSE and LON. Utilities is the highest component for both NYSE and LON.}
  \label{figure_4}
\end{figure}

Figure \ref{figure_1} shows that some eigenvalues are located outside the region predicted by Random Matrix Theory (eq. \ref{RMT_lambda}). These are the eigenvalues that we believe contain non-random information about the market \cite{{Laloux_PRL_83_1467_1999},{Plerou_PRL_83_1471_1999}}. We choose to study the three highest eigenvalues of each market and compare the results with each other. 
The eigenvector elements for the highest, $2^{nd}$ highest and $3^{rd}$ highest eigenvalues are represented in Figures \ref{figure_2}, \ref{figure_3} and \ref{figure_4}, respectively.

\begin{figure}[H]
  \centering
  \begin{tabular}{cc}
    \begin{minipage}{7.cm}
      \includegraphics[width=0.9\textwidth]{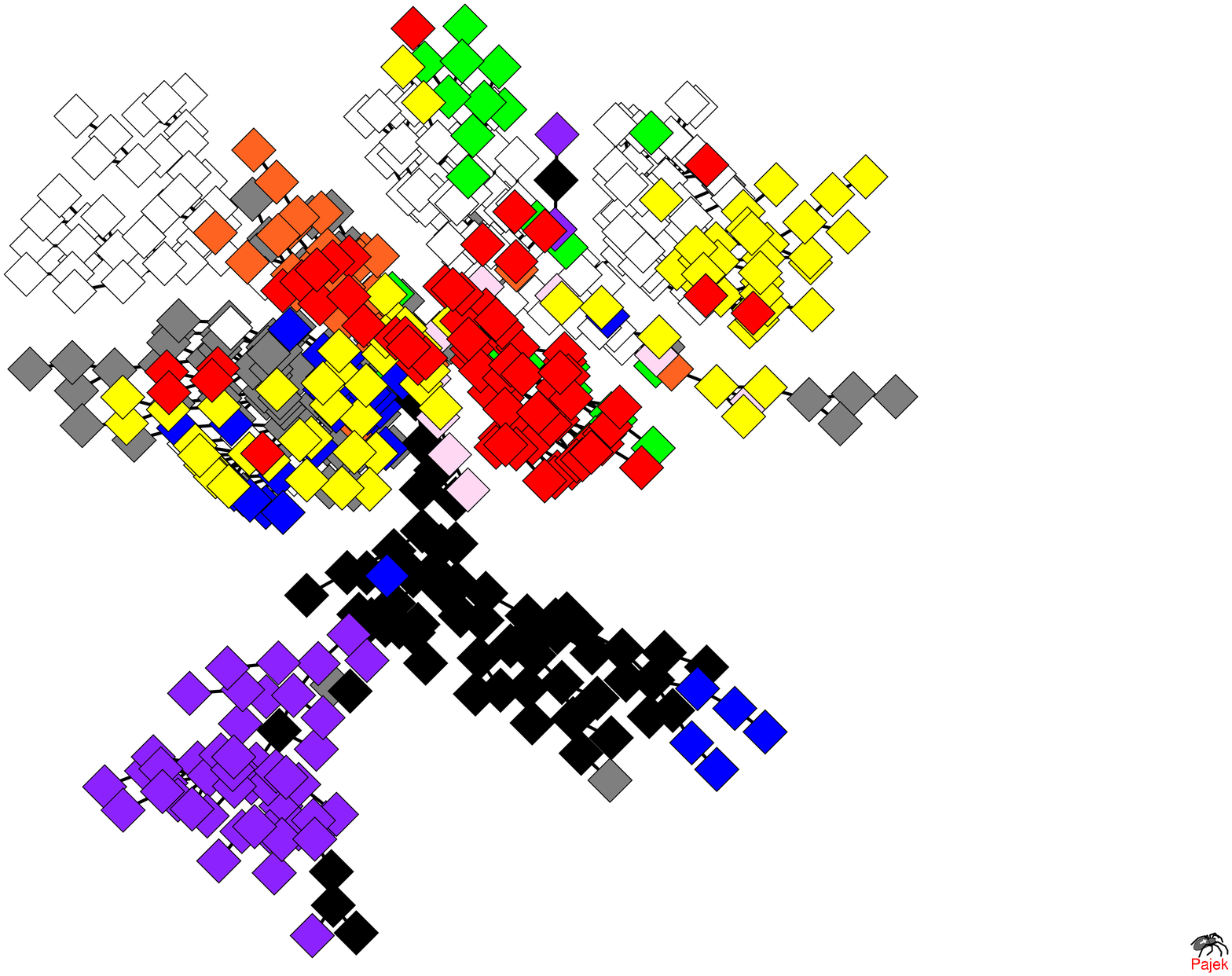}
      \begin{center}
	NYSE
      \end{center}
    \end{minipage}
    \begin{minipage}{7.cm}
      \includegraphics[width=0.9\textwidth]{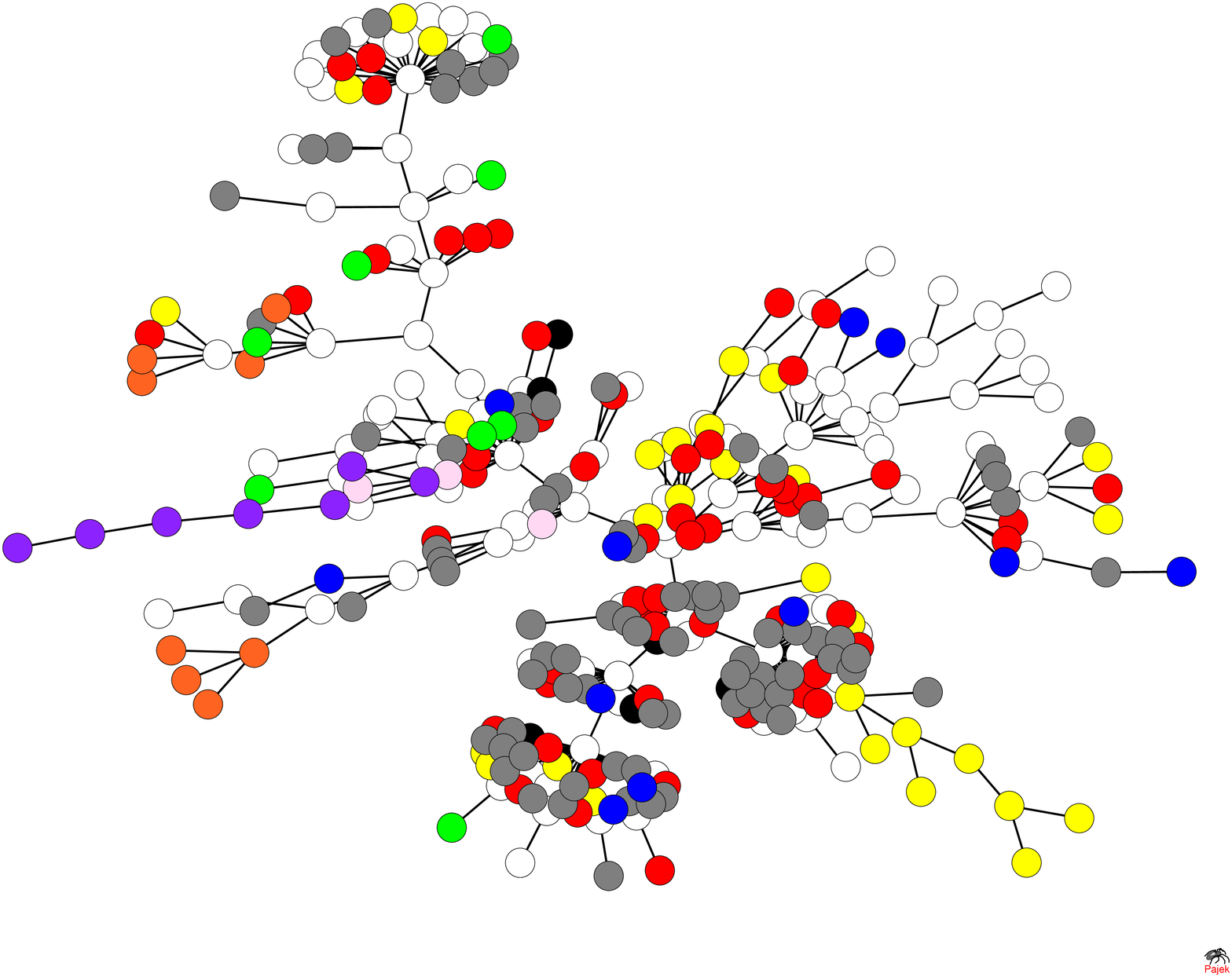}
      \begin{center}
	LON
      \end{center}
    \end{minipage}
  \end{tabular}
  \caption{Minimal Spanning Trees for two different markets: NYSE and LON. The colour code represents industrial sectors: black for oil and gas; blue for basic materials; grey for industrials; yellow for consumer goods; green for health care; red for consumer services; pink for telecommunications; purple for utilities; white for financials; orange for technology. NYSE stocks show clustering in industrial sectors while the LON tree shows a mixing of stocks from different industrial sectors.}
  \label{figure_5}
\end{figure}

Each eigenvector shows different industrial sectors that drive it. For example, as shown by other authors \cite{{Laloux_PRL_83_1467_1999},{Plerou_PRL_83_1471_1999},{Drozdz_PhysA_294_226_2001}}, for the eigenvector related with the highest eigenvalue all elements have the same sign, which means that all stocks contribute almost the same. This is known as the market mode and can be compared with the return index of the market that we are studying. For the eigenvector related with the $2^{nd}$ highest eigenvalue the stocks from different industrial sectors have different behaviours. 
For NYSE we see that the oil and gas and utilities sectors have the largest elements whereas in LON the two largest sectors are the technology and telecommunications sectors. In NYSE, the technology sector comes out third highest. Figure \ref{figure_4} shows the results for the $3^{rd}$ highest eigenvalue. Now we see that the utilities and technology sectors have the highest eigenvector components for both LON and NYSE. The third highest for NYSE is oil and gas whereas for LON it is the consumer goods sector.
Some of these strong sectorial correlations can be seen in Figure \ref{figure_5}, which shows the visualisation of the correlations between stocks using the MST. In the MST for NYSE these clusters are visible, however they are less obvious in the MST for LON.
As in our RMT analysis the sectors of oil and gas and utilities are singled out for NYSE. Here they feature as black and purple clusters at the bottom of the tree.
For the LON data, the situation is different: stocks from different industrial sectors are mixed together.

\section{Cross correlations between stocks of NYSE and LON}

\begin{figure}[H]
  \begin{center}
    \epsfysize=70mm
    \epsffile{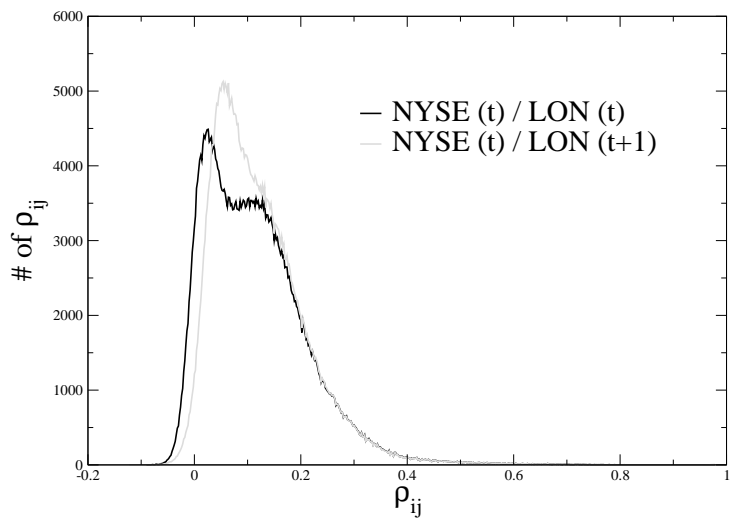}
    \caption{Distribution of the coefficients of the correlation matrix for the case of stocks from NYSE and LON at the same day (black line) and LON one day ahead of NYSE (grey line). The coefficients of the case where LON is one day ahead of NYSE are slightly more positive than in the previous case.}
    \label{figure_6}
  \end{center}
\end{figure}

Using the same techniques presented before, we also studied the cross correlations between stocks of different markets, in this case, NYSE and LON, for a portfolio of $939$ stocks from $10$ different industrial sectors. Because the data we use is the daily closing price of stocks, and we know that for example LON and NYSE close at different times, we also study the correlations between the stocks using the return of LON one day ahead of the return of NYSE. This results in a slight shift to the right of the distribution of coefficients from the correlation matrix (Figure \ref{figure_6}).

\begin{figure}[H]
  \begin{center}
    \epsfysize=70mm
    \epsffile{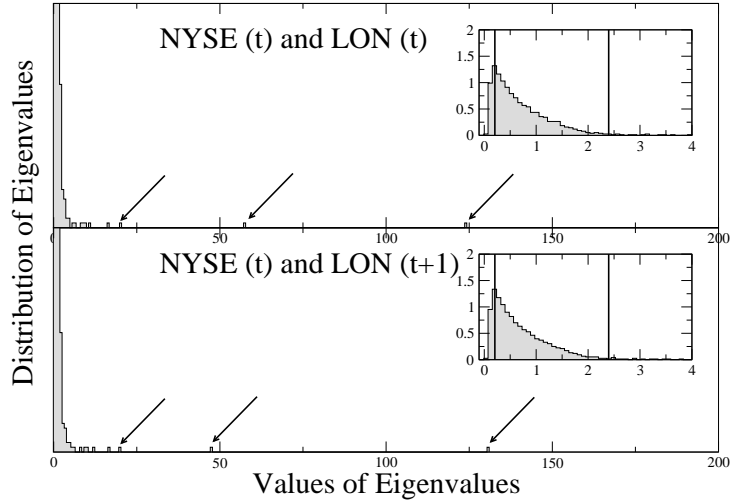}
    \caption{Spectrum of eigenvalues from the correlation matrices of cross correlations between stocks of NYSE and LON (top) and LON one day ahead of NYSE (bottom). The arrows show the three highest eigenvalues that we study more carefully. The vertical lines, in the inset figures, show the limits $\lambda_{min}^{max}$ (eq. \ref{RMT_lambda}).}
    \label{figure_7}
  \end{center}
\end{figure}

\begin{figure}[H]
  \centering
  \begin{tabular}{cc}
    \begin{minipage}{7.cm}
      \includegraphics[width=1.\textwidth]{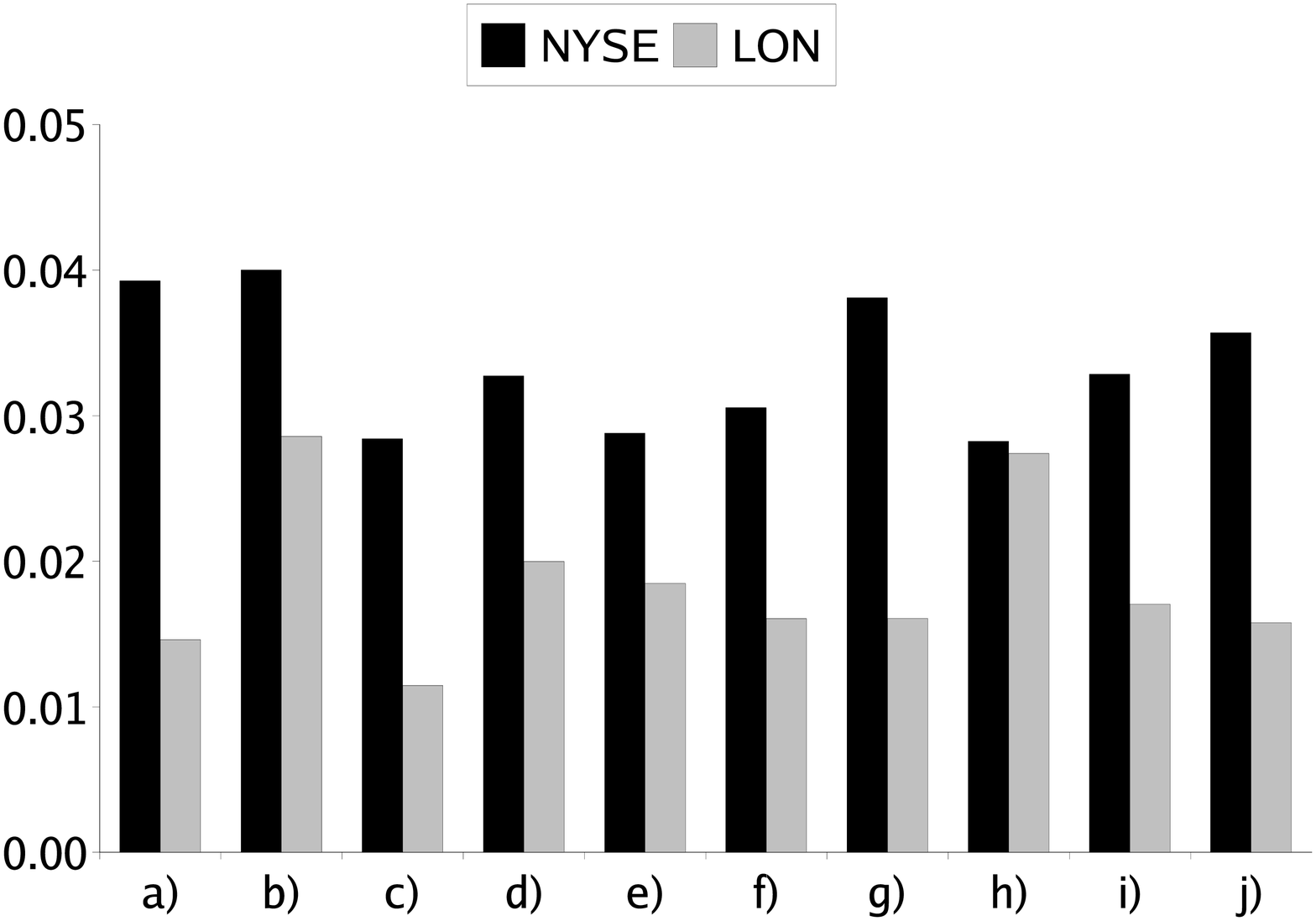}
      \begin{center}
	Highest
      \end{center}
    \end{minipage}
    \begin{minipage}{7.cm}
      \includegraphics[width=1.\textwidth]{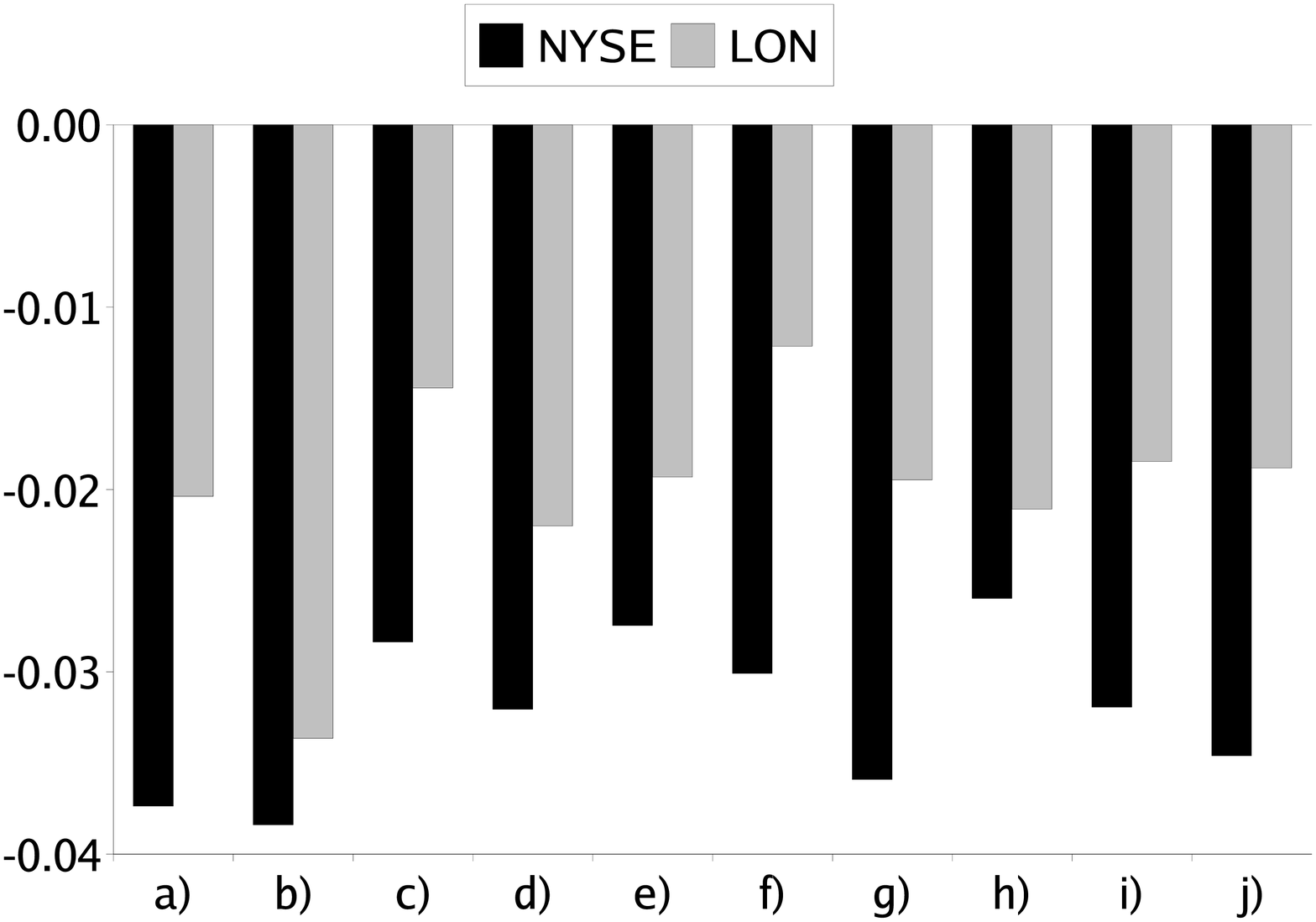}
      \begin{center}
	Highest
      \end{center}
    \end{minipage}
  \end{tabular}
  \begin{tabular}{cc}
    \begin{minipage}{7.cm}
      \includegraphics[width=1.\textwidth]{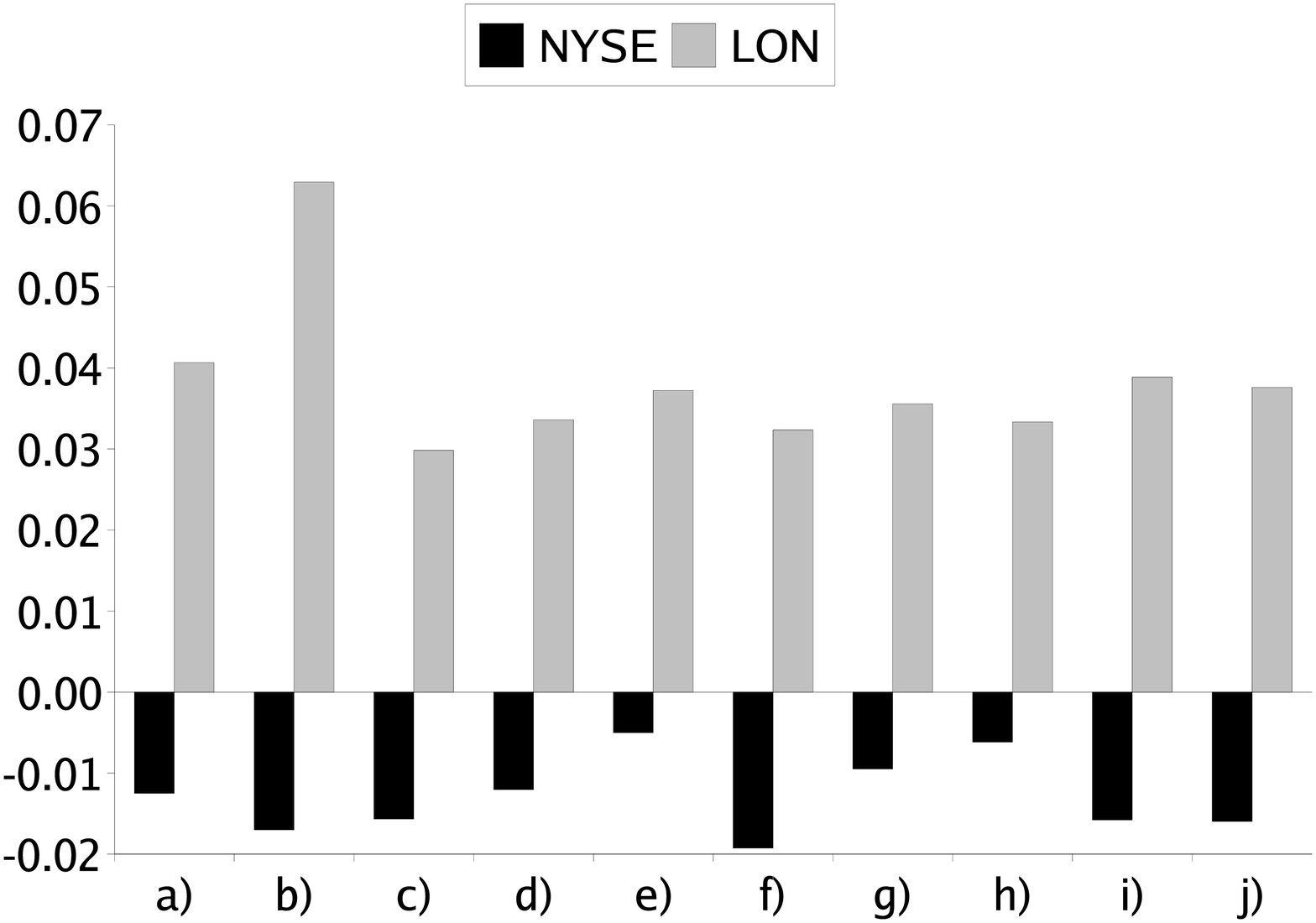}
      \begin{center}
	$2^{nd}$ highest
      \end{center}
    \end{minipage}
    \begin{minipage}{7.cm}
      \includegraphics[width=1.\textwidth]{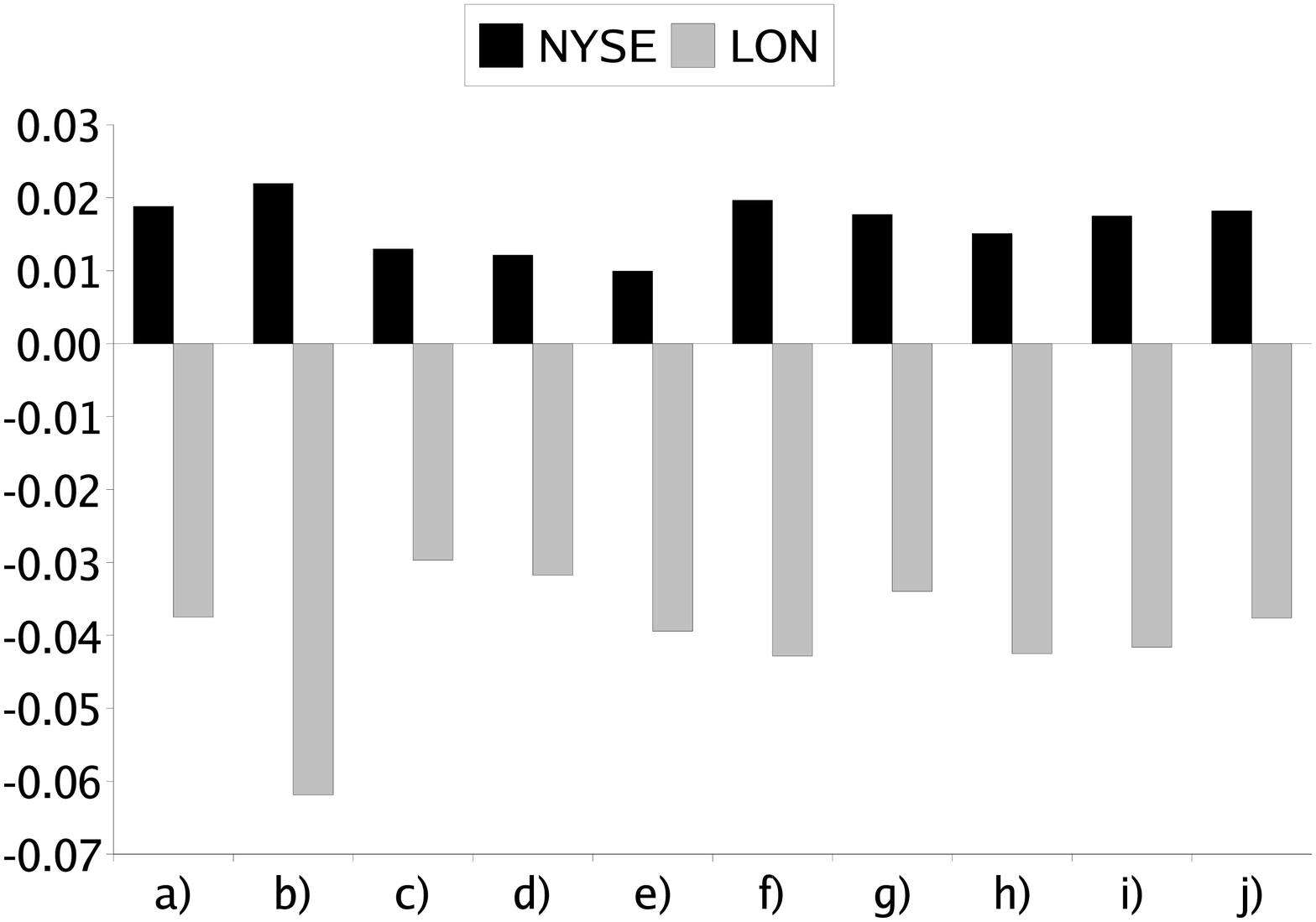}
      \begin{center}
	$2^{nd}$ highest
      \end{center}
    \end{minipage}
  \end{tabular}
  \begin{tabular}{cc}
    \begin{minipage}{7.cm}
      \includegraphics[width=1.\textwidth]{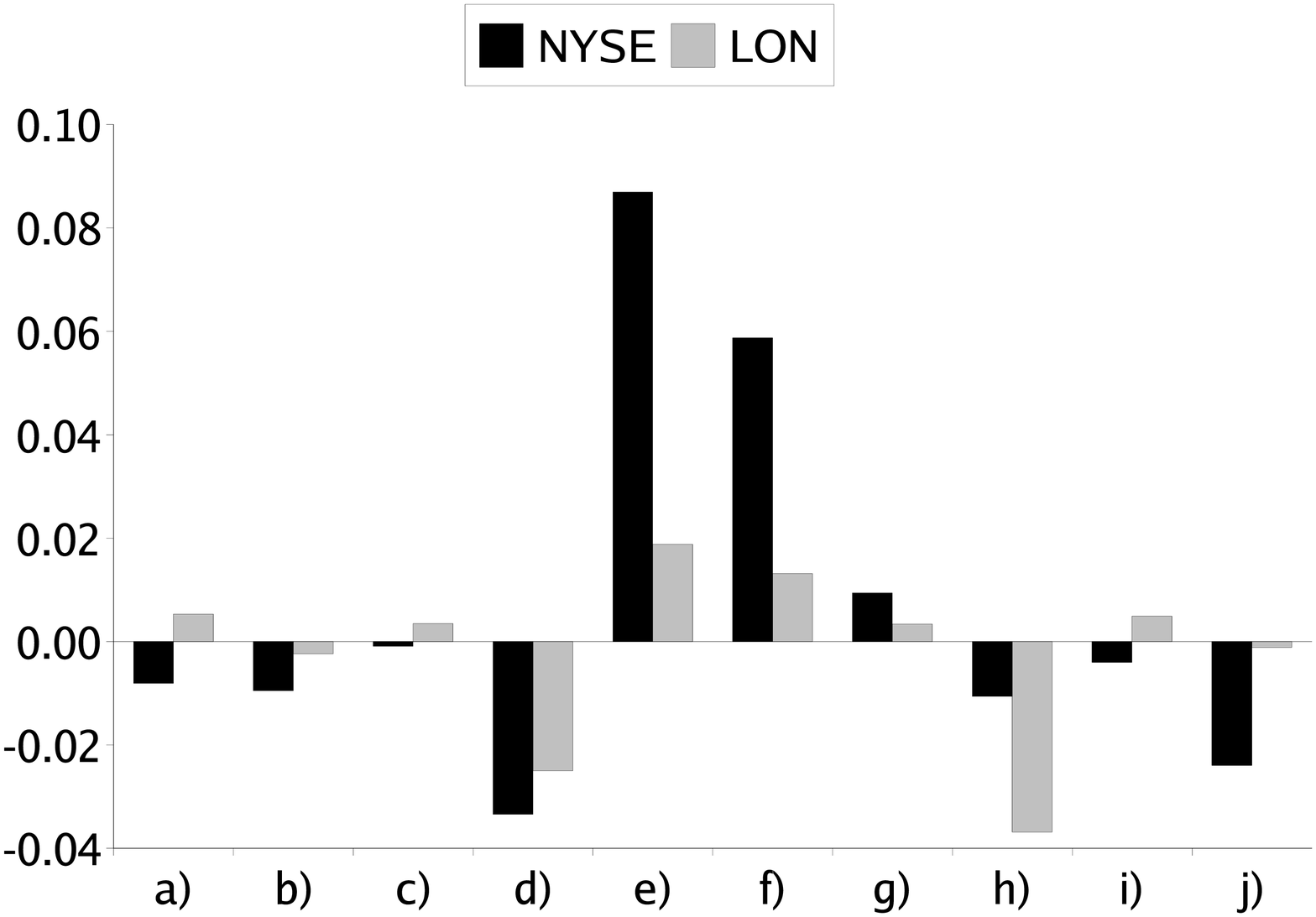}
      \begin{center}
	$3^{rd}$ highest
      \end{center}
    \end{minipage}
    \begin{minipage}{7.cm}
      \includegraphics[width=1.\textwidth]{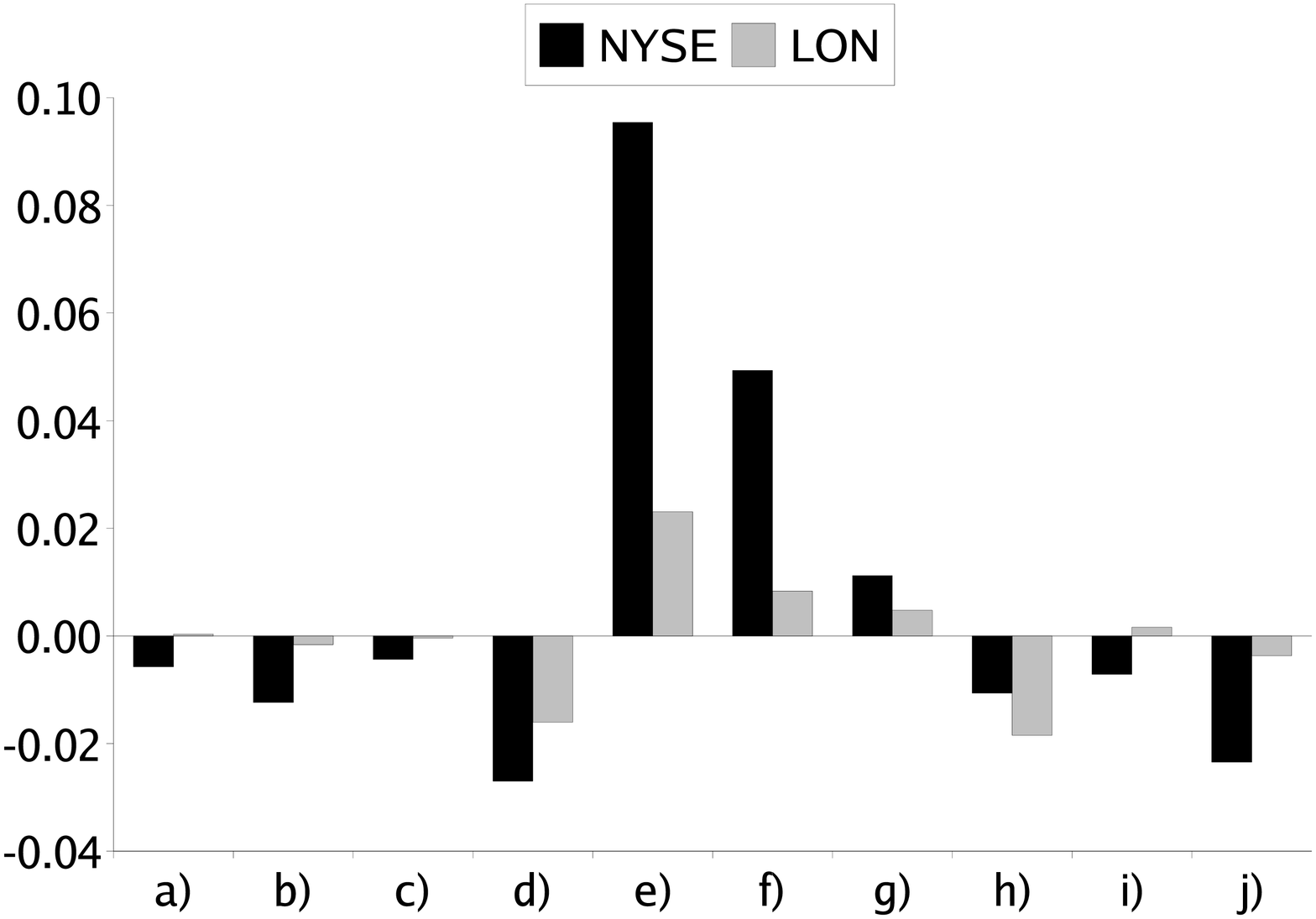}
      \begin{center}
	$3^{rd}$ highest
      \end{center}
    \end{minipage}
  \end{tabular}
  \caption{Eigenvectors elements of the highest, $2^{nd}$ and $3^{rd}$ highest eigenvalues for cross correlations between stocks of NYSE and LON (left) and LON one day ahead of NYSE (right). In the x axis we have the group of elements that belong to a industrial sector: a) industrials; b) financials; c) health care; d) technology; e) oil and gas; f) utilities; g) basic materials; h) telecommunications; i) consumer goods; j) consumer services. The black colour is for NYSE stocks and grey colour for LON stocks.}
  \label{figure_8}
\end{figure}

The distribution of eigenvalues of both correlation matrices can be seen in Figure \ref{figure_7}, where the highest eigenvalues are a mix of the highest eigenvalues of both markets for the individual studies.
The information contained in these eigenvalues show us how stocks from different markets are related to each other. Figure \ref{figure_8} shows the eigenvectors of the three highest eigenvalues.

\begin{figure}[H]
  \centering
  \begin{tabular}{cc}
    \begin{minipage}{7.cm}
      \includegraphics[width=0.9\textwidth]{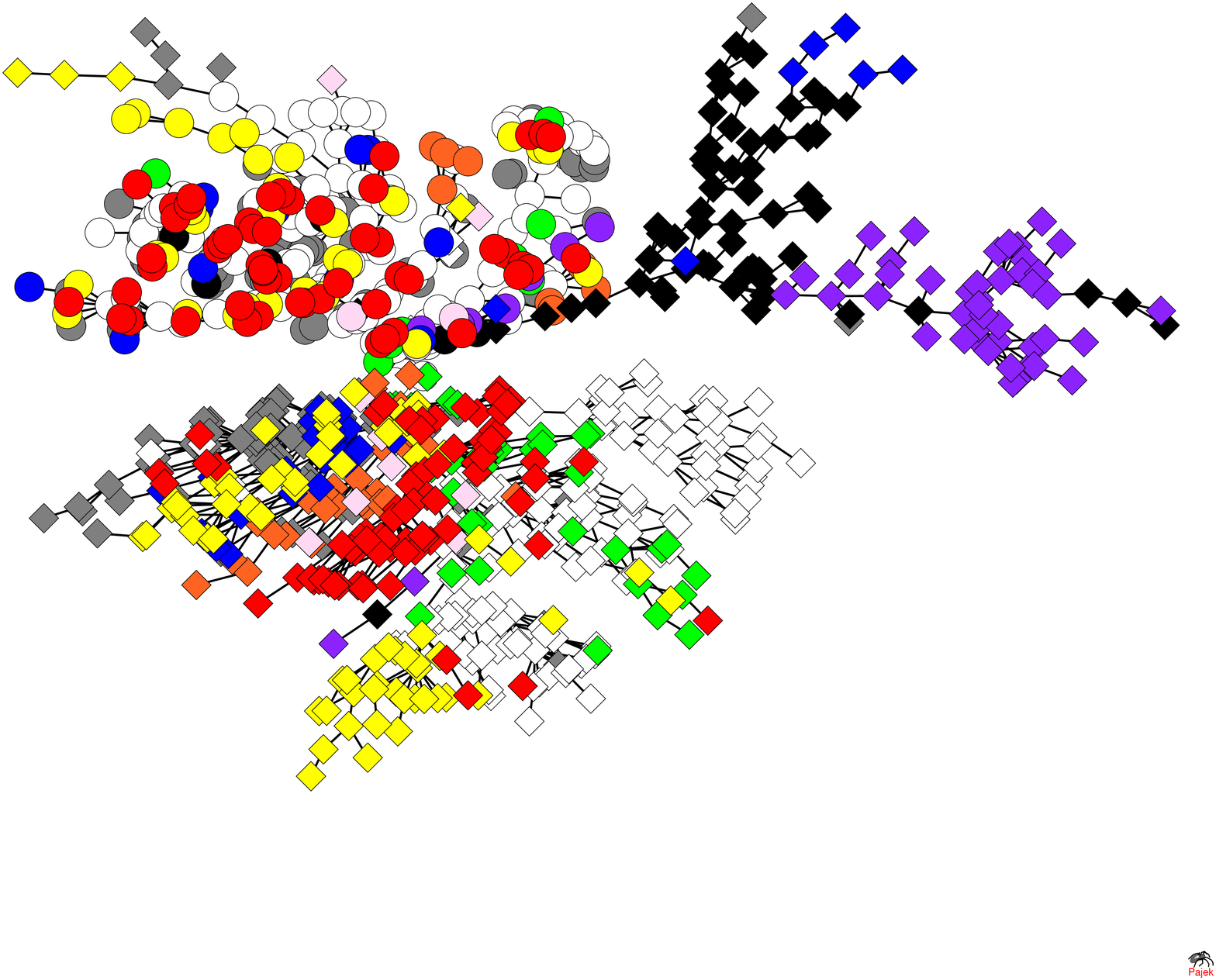}
      \begin{center}
	NYSE(t) and LON(t)
      \end{center}
    \end{minipage}
    \begin{minipage}{7.cm}
      \includegraphics[width=0.9\textwidth]{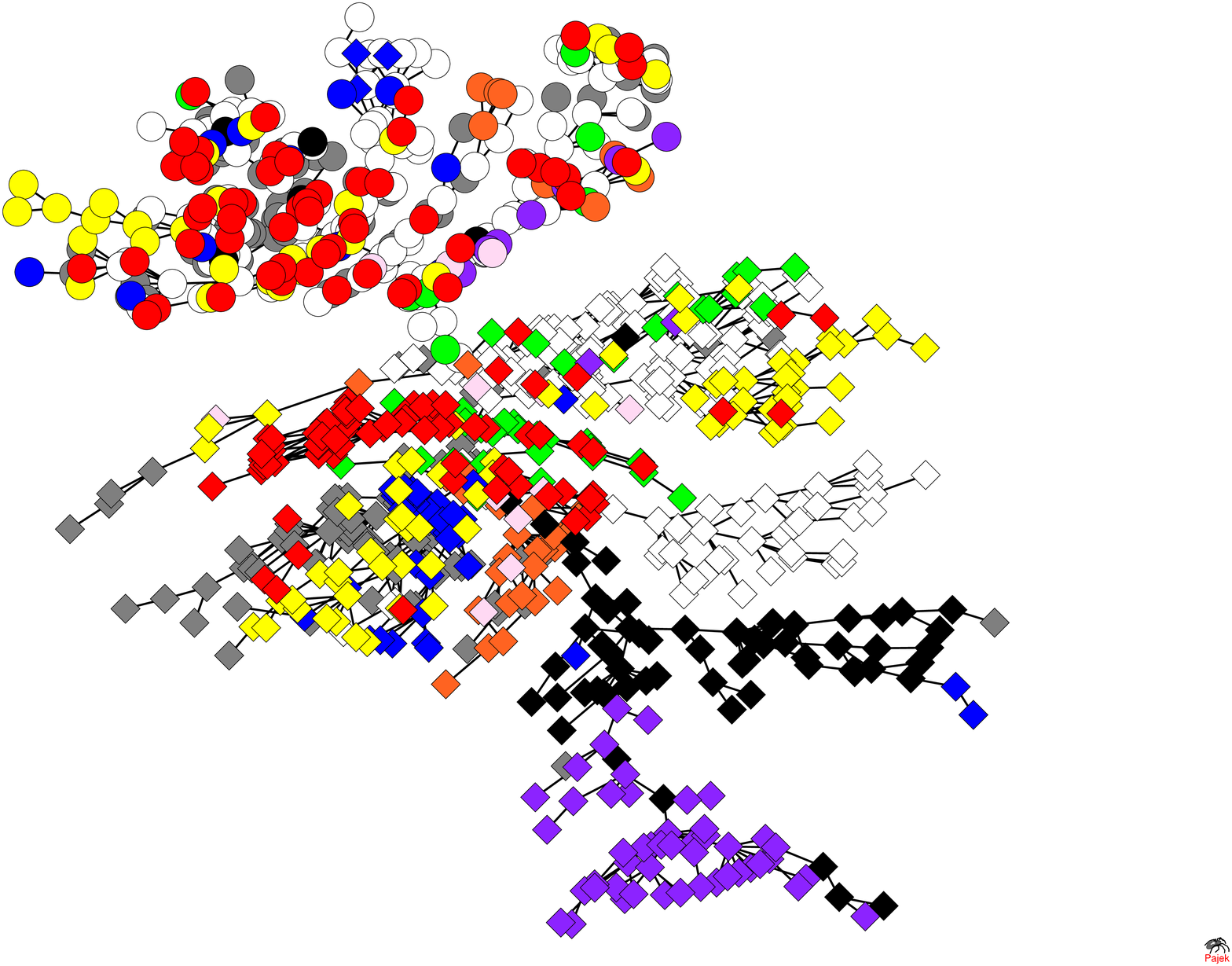}
      \begin{center}
	NYSE(t) and LON(t+1)
      \end{center}
    \end{minipage}
  \end{tabular}
  \caption{Minimal Spanning Tree for a portfolio of stocks from NYSE and LON markets with correlations computed at the same day (left) and when LON is one day ahead of NYSE (right). The colour code represents industrial sectors: black for oil and gas; blue for basic materials; grey for industrials; yellow for consumer goods; green for health care; red for consumer services; pink for telecommunications; purple for utilities; white for financials; orange for technology. The stocks from NYSE are represented by a lozenge ($\lozenge$) and the stocks from LON are represented by a circle ($\circ$).}
    \label{figure_9}
\end{figure}

The eigenvector related to the highest eigenvalue shows that all the stocks from different markets and sectors follow the same trend (market mode), just as in the study of the individual markets (Figure \ref{figure_2}). For this portfolio of stocks, the markets remain separated as before as is evident from the results for the $2^{nd}$ largest eigenvalue. The eigenvector related to the $3^{rd}$ highest eigenvalue shows what we saw for the $2^{nd}$ highest eigenvalue of the NYSE market, with a bigger influence of oil and gas and utilities sectors (Figure \ref{figure_3}). For the sectors of LON the comparison with the $2^{nd}$ highest eigenvalue of the individual study is not so clear. In the case where the correlations were computed at the same day telecommunications and technology continue to have a bigger influence (Figure \ref{figure_3}), but oil and gas and utilities also have a bigger influence in this eigenvector, probably pulled by the fact that these are the sectors of NYSE that influence this eigenvector. In the case where LON is one day ahead of NYSE, this influence is even more clear, with oil and gas to be the sector with a bigger influence in this eigenvector. So we can see that NYSE has pulled the LON market more into line with NYSE. This is not so easily seen in the MST (Figure \ref{figure_9}) that simply continues to show the geographical separation of the markets as reflected in the data for the $2^{nd}$ highest eigenvector.
Note that these results are essentially unchanged whether we evaluate the correlations on the same day (where the closing of NYSE is after that of LON) or whether we evaluate the correlations using for LON the day after that of NYSE (essentially testing whether LON follows NYSE). The main change in the MST is the rather curious shift in the position of the oil and gas and utilities sectors.

\section{Conclusion}
We have used two different methods to study correlations between London and New York stocks. Our results using Random Matrix Theory show that the markets remain largely separate even when cross correlations between stocks across the two markets are included. Only at the level of the $3^{rd}$ highest eigenvalue are significant changes seen and these take the form of New York effectively modifying the London positions with the New York data remains broadly unchanged. The results for the Minimal Spanning Trees broadly reflect the results from the Random Matrix Theory. But it is not as easy to see the detail provided by the Random Matrix analysis. This of course is not too surprising since the Minimal Spanning Trees approach only uses partial information from the correlation matrix.

Much finance research has addressed the issue of whether or not stocks ultimately cluster by market or by industry. There is no consensus on this. Some \cite{Flavin_JIMF_23_1137_2004} suggest that the clustering is primarily industrial, while others \cite{Phylaktis_EJF_12_627_2006} contend that the split is primarily geographical. The evidence here is that geographical (more correctly, market) location is the most important element in determining the cluster into which a stock falls. The implication for portfolio managers is that, at least at a first level, they should consider diversification along market lines, and only subsequently along industrial or sectoral lines

Further research on this approach is very possible. An obvious extension is to examine the market dynamics, as revealed by clustering, of stocks that share the market. Two types of sharing are possible: stocks can be cross-listed, with a listing on both markets, or they can be listed via the issuance of depository receipts. If there are truly different dynamics at work in the two markets then these stocks provide a natural experiment to investigate this. A further expansion would be to examine whether these clusters here prevail if we consider unhedged investors, examining the stocks in the currency of the market country.

\begin{ack}
R. Coelho acknowledges the support of the FCT/Portugal through the grant SFRH/BD/$27801/2006$. P. Richmond acknowledges help from COST action P10 Physics of Risk. B. Lucey acknowledges support from Irish Government via the Programme for Research in Third Level Institutions.
\end{ack}


\begin{thebibliography}{99}
%
\bibitem{Mantegna_EPJB11_193_1999} R. N. Mantegna, {\it Hierarchical structure in financial markets}, Eur. Phys. J. B \textbf{11}, 193 (1999)
%
\bibitem{Onnela_PRE68_056110_2003} J.-P. Onnela, A. Chakraborti, K. Kaski, J. Kert\'esz and A. Kanto, {\it Dynamics of market correlations: Taxonomy and portfolio analysis}, Phys. Rev. E \textbf{68}, 056110 (2003)
%
\bibitem{Coelho_PhysA_373_615} R. Coelho, S. Hutzler, P. Repetowicz and P. Richmond, {\it Sector analysis for a FTSE portfolio of stocks}, Physica A 373 (2007) 615-626 
%
\bibitem{Bonanno_PRE62_7615_2000} G. Bonanno, N. Vandewalle and R. N. Mantegna, {\it Taxonomy of stock market indices}, Phys. Rev. E \textbf{62}, 7615 (2000)
%
\bibitem{Coelho_PhysA_376_455} Ricardo Coelho, Claire G. Gilmore, Brian Lucey, Peter Richmond and Stefan Hutzler, {\it The Evolution of Interdependence in World Equity Markets - Evidence from Minimum Spanning Trees}, Physica A 376 (2007) 455-466.
%
\bibitem{Gopikrishnan_PRE_64_035106_2001} P. Gopikrishnan, B. Rosenow, V. Plerou and H. E. Stanley, {\it Quantifying and interpreting collective behavior in financial markets}, Phys. Rev. E \textbf{64}, 035106 (2001)
%
\bibitem{Plerou_PRE_65_066126_2002} V. Plerou, P. Gopikrishnan, B. Rosenow, L. A. N. Amaral, T. Guhr and H. E. Stanley, {\it Random matrix approach to cross correlations in financial data}, Phys. Rev. E \textbf{65}, 066126 (2002)
%
\bibitem{Utsugi_PRE_70_026110_2004} A. Utsugi, K. Ino and M. Oshikawa, {\it Random matrix theory analysis of cross correlations in financial markets}, Phys. Rev. E \textbf{70}, 026110 (2004)
%
\bibitem{Kim_PRE72_046133_2005} D.-H. Kim and H. Jeong, {\it Systematic analysis of group identification in stock markets}, Phys. Rev. E \textbf{72}, 046133 (2005)
%
\bibitem{ICBClassification}http://www.icbenchmark.com/
%
\bibitem{Mantegna_Book} R. N. Mantegna and H. E. Stanley, {\it An Introduction to Econophysics: Correlations and Complexity in Finance.} Cambridge University Press, Cambridge (2001)
%
\bibitem{Sengupta_PRE_60_3389} A. M. Sengupta and P. P. Mitra, {\it Distributions of singular values for some random matrices}, Phys. Rev. E \textbf{60}, 3389 (1999)
%
\bibitem{Prim} R. C. Prim, {\it Shortest connection networks and some generalisations}, Bell System Tech. J. \textbf{36}, 1389 (1957)
%
\bibitem{Laloux_PRL_83_1467_1999} L. Laloux, P. Cizeau, J.-P. Bouchaud and M. Potters, {\it Noise Dressing of Financial Correlation Matrices}, Phys. Rev. Lett. \textbf{83}, 1467 (1999)
%
\bibitem{Plerou_PRL_83_1471_1999} V. Plerou, P. Gopikrishnan, B. Rosenow, L. A. N. Amaral and H. E. Stanley, {\it Universal and Nonuniversal Properties of Cross Correlations in Financial Time Series}, Phys. Rev. Lett. \textbf{83}, 1471 (1999)
%
\bibitem{Drozdz_PhysA_294_226_2001} S. Dro\.zd\.z, F. Gr\"ummer, F. Ruf and J. Speth, {\it Towards identifying the world stock market cross-correlations: DAX versus Dow Jones}, Physica A \textbf{294}, 226 (2001)
%
\bibitem{Flavin_JIMF_23_1137_2004} T. J. Flavin, {\it The effect of the Euro on country versus industry portfolio diversification}, Journal of International Money and Finance \textbf{23}, 1137 (2004)
%
\bibitem{Phylaktis_EJF_12_627_2006} K. Phylaktis and L. Xia, {\it The Changing Roles of Industry and Country Effects in the Global Equity Markets}, The European Journal of Finance \textbf{12}, 627 (2006)
%
\end{thebibliography}
\end{document}